%% file: Main.tex
\documentclass{article}

\usepackage{arxiv}

\usepackage{array}
\usepackage{float}
\usepackage{graphicx}
\usepackage{nomencl}
\usepackage{multirow}
\usepackage{dirtytalk}
\usepackage{tabto}
\usepackage{comment}
\usepackage{subcaption}
\usepackage{adjustbox}
\usepackage{makecell}
\usepackage{longtable}
\usepackage{mathtools}
\usepackage{array,ragged2e}
\usepackage{booktabs}
\usepackage{cite}
\usepackage{tabularx}
\usepackage{xcolor}
\usepackage{ifthen}
\usepackage{rotating}
\usepackage{siunitx}
\usepackage{vcell}
\usepackage{hyperref}
\usepackage{arydshln}
\usepackage{color, colortbl}
\usepackage{relsize}
\usepackage{placeins}
\usepackage{multirow}%
\usepackage{amsmath,amssymb,amsfonts}%
\usepackage{amsthm}%
\usepackage{mathrsfs}%
\usepackage[title]{appendix}%
\usepackage{textcomp}%
\usepackage{manyfoot}%
\usepackage{algorithm}%
\usepackage{algorithmicx}%
\usepackage{algpseudocode}%
\usepackage{listings}%

\title{CVE-driven Attack Technique Prediction with Semantic Information Extraction and a Domain-specific Language Model}

\author{ {Ehsan Aghaei} \\
	Department of Computer Science\\
	Carnegie Mellon University\\
	Pittsburgh, PA \\
	\texttt{eaghaei@andrew.cmu.edu} \\
	\And
	{Ehab Al-Shaer} \\
	Department of Computer Science\\
	Carnegie Mellon University\\
	Pittsburgh, PA \\
	\texttt{ealshaer@andrew.cmu.edu} \\
}
\date{}

\begin{document}

\maketitle

\begin{abstract}
This paper addresses the critical need for bridging the gap between vulnerability information, as represented by Common Vulnerabilities and Exposures (CVEs), and the resulting attack actions. While CVEs offer insights into security vulnerabilities and their exploitations, they often lack the intricate particulars that point to potential threat actions denoted by tactics, techniques, and procedures (TTPs) within the ATT\&CK framework. This poses considerable challenges for cybersecurity practitioners and developers in accurately categorizing CVEs based on their potential threats and proactively initiating countermeasures against attacks.

We present novel techniques implemented in the TTPpredictor tool to analyze the CVE description and infer the plausible TTP attacks caused by exploiting this CVE. The realization of TTPpredictor hinges on addressing of two key challenges. First, the scarcity of well-labeled datasets suitable for training makes the task of classifying CVEs into TTPs unfeasible. Second, the disparity in semantics between CVE and TTP descriptions within ATT\&CK renders the process of mapping and automating labeling exceptionally intricate.

To tackle these challenges, we initially extracted a substantial volume of threat actions from an extensive collection of unstructured cyber threat reports, encompassing CVEs, through the application of Semantic Role Labeling (SRL) techniques. Subsequently, we correlated these threat actions, along with their contextual attributes such as manner and purpose, with the attack functionality classes delineated by MITRE by simply matching the verb-object (VO). 
As a result, this automated correlation between the new context of threat actions and their corresponding threat functionalities facilitates the process of creating automatically labeled data. This data becomes instrumental in categorizing novel threat actions (VOs) into both threat functionality classes and TTPs.

Through an iterative refinement procedure over a considerable corpus of compiled threat action sentences, we constructed a comprehensive dataset that serves as a pivotal asset for training an accurate TTPpredictor model. Subsequently, we utilized this expansive annotated dataset to fine-tune our domain-specific language model for cybersecurity, termed SecureBERT. The extended model, TTPpredictor, captures both paradigmatic relationships among threat actions and meticulous semantic associations between CVE threat actions, their context, and attack techniques. This integration of novel techniques leads to a precise and dependable TTPpredictor.

An empirical assessment of our approach showcases its effectiveness, achieving notable accuracy rates of approximately 98\% and F1-scores ranging from 95\% to 98\% in the precise classification of CVEs to their corresponding ATT\&CK techniques. Furthermore, we demonstrate that the performance of the TTPpredictor outperforms that of state-of-the-art language model tools such as ChatGPT.
\end{abstract}


\input{Introduction.tex}
\input{Problem_Definition.tex}

\input{DataAssessment.tex}
\input{Model_Design.tex}
\input{Evaluation.tex}

\input{RelatedWorks.tex}
\input{Conclusion.tex}

\bibliographystyle{IEEEtranS.bst}  
\bibliography{sn-bibliography.bib}

\end{document}

%% file: Introduction.tex
\section{Introduction}
Classifying Common Vulnerabilities and Exposures (CVEs) to Tactics, Techniques, and Procedures (TTPs) is a crucial task in cybersecurity. This classification provides valuable insights into the specific techniques used by threat actors to exploit vulnerabilities. However, challenges such as the lack of labeled datasets, semantic gaps, and the absence of domain-specific resources make this classification process challenging. To address these challenges, this paper introduces the use of SecureBERT \cite{aghaei2022securebert}, a domain-specific language model, along with semantic role labeling (SRL)~\cite{he2017deep} for data collection.

CVEs represent identified vulnerabilities, while TTPs encompass the tactics and techniques used by threat actors. Understanding the relationship between CVEs and TTPs helps security analysts assess the severity and impact of vulnerabilities accurately. By classifying CVEs to commonly exploited TTPs, analysts can gain insights into the specific techniques employed by threat actors. This classification of CVEs to TTPs also intersects with the field of cyber threat intelligence (CTI). CTI involves gathering and analyzing information about potential threats and adversaries to inform defensive strategies. 

Mapping CVEs to TTPs enriches organizations' understanding of threat actors' tactics and techniques. This knowledge enables proactive threat identification, prioritization of defense measures, and resource allocation. By characterizing CVEs in the context of TTPs, organizations can identify patterns in attack behavior and develop robust defense strategies.
However, the classification of CVEs to  TTPs presents several challenges. One significant challenge is the lack of a labeled dataset that directly maps CVEs to their corresponding TTPs. This scarcity of data poses challenges in developing and evaluating accurate classification models. Furthermore, there exists a semantic gap between CVEs, which are typically described using technical language specific to vulnerabilities, and TTPs, which encompass a broader range of tactics and techniques. Bridging this semantic gap requires a robust approach that captures the nuanced relationships and context between CVEs and TTPs.

To overcome these challenges, this paper presents a novel model that characterizes the threat actions based on its context using our domain-specific language model, SecureBERT, for cybersecurity. This semantic characterization, along with semantic role labeling techniques, aims to automate the classification of CVEs to the most potent attack techniques as stated in the ATT\&CK framework through threat functionalities. Threat functionalities represent the specific capabilities acquired by attackers after exploiting a vulnerability. These "Functionalities" are then mapped to the most commonly used attack techniques, enabling the classification of CVEs to both functionalities and techniques~\cite{MITREFUNC}.
 
 Our approach employs Subject-Verb-Object (SVO) representations to extract functionalities from security reports, like CVEs. Threat functionalities are captured using manually annotated dictionaries and Semantic Role Labeling (SRL) techniques. Extracted SVOs provide a concise depiction of threats and highlight relationships between them. The model for capturing linguistic connections uses SVO structures and additional pairs for learning similarities, compensating for the lack of labeled training data. Specifically, the extracted SVOs provide a simple representation of the threats while identifying the paradigmatic relation between different SVOs, and the semantic dependencies between the SVOs and the surrounding context considering the broader context of the security report, enhance the understanding of the intended threat.

Capturing the paradigmatic relationship between various SVOs is pivotal for expanding the representations of threat actions. For instance, the connection between "attacker reads files" and "attacker opens file" lies in their shared semantic concept of interacting with files for information retrieval. Although these phrases share commonalities, distinctions in specific conveyed actions can arise. Both might involve accessing a file, and potentially reading its content. Yet, one might allude to unauthorized data access, while the other could relate to data exfiltration or reconnaissance. Alternatively, they might denote an action to exploit or define an attack's impact.

Comprehending context is vital for precise interpretation and effective threat analysis. Context furnishes supplementary information that aids in understanding the intentions and implications of actions. For instance, in the mentioned scenario, context can elucidate whether the attacker intends unauthorized access or file system exploitation. Ignoring context analysis may lead to misinterpretation, misclassification, and inadequate security measures. Hence, considering broader context alongside paradigmatic relations enhances threat detection accuracy, enabling proactive risk mitigation.

The importance of integrating content and context as inputs amplifies due to the interdependent nature of functionalities. Identifying subject-verb-object (SVO) structures heavily relies on contextual data. In complex cases, multiple functionalities share the same SVO, demanding context for precise interpretation. Take "attackers read memory" as an example. Its meaning varies based on context. In the context of "Attackers read memory and cause a denial of service," it signifies memory reading as the attack's impact. Yet, in the context of "Attackers read memory to discover Username and Password," it denotes the action leading to credential theft. Therefore, using context alone as input is unfeasible, as it often involves various threat-related statements. The extraction process captures functionality-related SVOs without knowledge of other potential functionality representations in the same report. Thus, accurately labeling reports becomes challenging. Integrating both content (SVO) and context as inputs is crucial for the classification model, ensuring accurate and comprehensive functionality categorization.

By seamlessly integrating the extracted SVOs with their contextual surroundings, the presented approach endeavors to authentically portray functionalities, even in the absence of abundant ground truth data. This methodology enables the identification of intricate actions and latent patterns that might not be overtly articulated within the content itself.

This paper's structure unfolds as follows. We first describe the threat functionality classes and the challenges of this work in Section~\ref{sec: functionality1} and Section~\ref{sec: problem_def} in Section~\ref{sec: data_collection}, respectively. The data extraction, annotation, and assessment are described in Section~\ref{sec: data_collection}. Finally, Section~5 and Section~\ref{sec: eval} describe our CVE-to-TTP classification and the evaluation results, respectively.  

\section{Functionality}\label{sec: functionality1}
Functionalities are categorized into two groups: action functionalities and impact functionalities. Action functionalities describe specific actions taken by attackers, such as reading files or creating accounts, while impact functionalities focus on the consequences or outcomes of the attack, such as memory errors.
The MITRE framework presents a set of sixteen functionalities (see Table \ref{tab: functionality_list}) that describe specific actions an attacker can take when exploiting a vulnerability. However, accurately classifying CVEs to their corresponding attack functionalities presents several challenges. The most significant challenge is the lack of labeled data specifically mapping CVEs to attack functionalities.
To address this, it is necessary to manually collect and annotate a standardized dataset to train the predictive model. Nonetheless, data collection and annotation present their own challenges, as they require comprehensive text analysis to understand the language structure of CVEs and extract relevant text patterns for creating a concise dataset.
\begin{table}
\small
\centering
\begin{tabular}{|l|l|l|} 
\hline
\multicolumn{3}{|c|}{\textbf{Common Functionalities}}                   \\ 
\hline \hline
Modify
  Configuration & Create
  Account                           & Disable
  Protections              \\ 
\hline
Restart/Reboot         & Install
  App                              & Read
  from Memory                 \\ 
\hline
\multicolumn{2}{|l|}{Obtain
  Sensitive Information: Credentials}   & Password
  Reset                   \\ 
\hline
\multicolumn{2}{|l|}{Obtain
  Sensitive Information: Other Data}    & Read
  Files                       \\ 
\hline
Delete
  Files         & Create/Upload
  File                       & Write
  to Existing File           \\ 
\hline
\multicolumn{3}{|l|}{Change
  Ownership or Permissions}                 \\ 
\hline
\multicolumn{3}{|l|}{Memory Modification (Memory Buffer
  Errors, Pointer Issues, Type Errors, etc.)}    \\ 
\hline
\multicolumn{3}{|l|}{Memory Read (Memory Buffer Errors,
  Pointer Issues, Type Errors, etc.)}            \\
\hline

\end{tabular}
\caption{List of common functionalities defined by MITRE. A "Functionality" represents an ability the attacker gains access to after exploiting an attack.}
\label{tab: functionality_list}
\end{table}

Furthermore, attack functionalities are not entirely independent, as they may exhibit mutual characteristics or dependencies. For instance, in the language of cybersecurity, the action of an attacker "reading a file" can imply "obtaining sensitive information." CVE descriptions often convey such capabilities, either explicitly or implicitly. For example, descriptions such as "\textit{attacker uninstalled antivirus software}" and "\textit{attacker compromised the firewall's functionality}" both indicate manipulation of protection systems to hinder their effectiveness. Therefore, studying CVE descriptions alone can provide valuable information in identifying these functionalities. Recognizing these functionalities on the defensive side contributes to improved vulnerability management. Similar to vulnerability types, functionalities are also associated with a range of MITRE techniques, enabling cybersecurity researchers to articulate the exploitation process of a vulnerability more effectively. This, in turn, enhances cyber threat intelligence and facilitates more robust defense planning.
\newline
\\
\noindent{\textbf{Functionality Documentation}}\\
MITRE guideline has provided sixteen functionality names \cite{MITREFUNC} without any particular definition.
hence, we define sixteen different most common functions an attacker may be trying to gain access to through the exploitation known as functionalities as follows:

If $f_z$ denotes the functionality $f$ with index $z$, let's define the functionalities described in Table \ref{Tab: func_desc}.

\begin{figure}[!ht]
 \center
  \includegraphics[width=14cm]{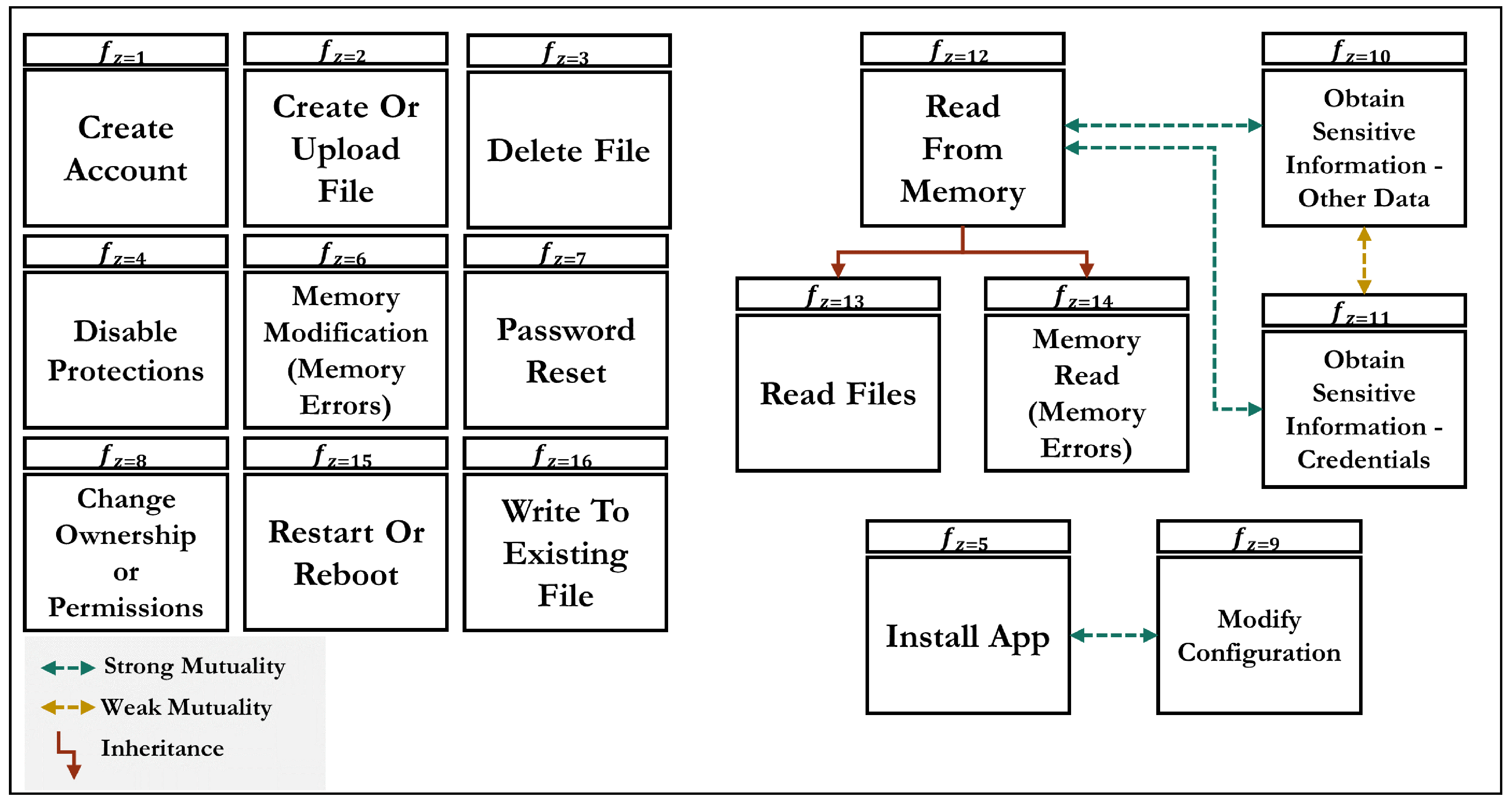}
  \caption{This figure demonstrates the functionalities and their different types of relationships.}
  \label{fig: functionality_relat}
\end{figure}

\begin{table*}[!ht]
\vspace{20pt}
\scriptsize
\centering
\begin{tabular}{|p{4cm}|p{8cm}|} 
\hline
\vspace{0.2pt}
\textbf{Functionality Name (ID)} & \textbf{Description}\\
\hline
\vspace{0.2pt}
\textbf{Create Account ($f_{z=1}$)} & the act of unauthorized creation of new accounts or adding new users to the victim system done by the attacker.\\ \hline
\vspace{0.2pt}
\textbf{Create Or Upload File ($f_{z=2}$)} & the act of unauthorized creation or uploading any file to any system for any purpose done by the attacker.\\ \hline
\vspace{0.2pt}
\textbf{ Delete Files ($f_{z=3}$)} & the act of unauthorized deletion or destruction of any information including but not limited to files, contents, data, etc., done by the attacker. It is different from data manipulation.
\\ \hline
\vspace{0.2pt}
\textbf{Disable Protections ($f_{z=4}$)} & the act of causing any malfunction, interruption, or abnormality in any security/defensive process or system such as anti-viruses, anti-malware, authentication procedures, firewall, security checks, etc., done by the attacker.
\\ \hline
\vspace{0.2pt}
\textbf{Install App ($f_{z=5}$)} & the act of delivering and/or installing any malicious application or configuration on the victim's system causing further threats, done by the attacker.
\\ \hline
\vspace{0.2pt}
\textbf{Memory Modification (Memory Buffer Errors, Pointer Issues, Type Errors, etc.) ($f_{z=6}$)} & the act of any invalid modification, manipulation, and/or write to the memory (e.g., buffer, kernel, memory locations, pointers, etc.) leading to memory issues such as buffer-over read, memory crash, buffer overflow, etc., done by the attacker.
\\ \hline
\vspace{0.2pt}
\textbf{Password Reset ($f_{z=7}$)} & the act of manipulating an account such as modifying credentials (e.g., ID, username, password, email account name, etc.) for any purpose, done by the attacker.
\\ \hline
\vspace{0.2pt}
\textbf{Change Ownership or Permissions ($f_{z=8}$)} & the act of changing file ownership, and/or modifying access permission (access controls) for any purpose, done by the attacker.
\\ \hline
\vspace{0.2pt}
\textbf{Modify Configuration ($f_{z=9}$)} & the act of modifying, editing, and manipulating any systems configuration and/or settings causing further threats, done by the attacker.
\\ \hline
\vspace{0.2pt}
 & \textit{NOTE 1} The actions and intention in "Install App" and "Modify Configuration" are quite similar sharing similar techniques (not same), and since there is no ground truth available to distinguish them, we combine these two and considered them as a single functionality.\\
\hline
\vspace{0.2pt}
 \textbf{Obtain Sensitive Information - Other Data ($f_{z=10}$)} & the act of obtaining any non-credential sensitive information without authorization via any method (unauthorized access to files, databases, memory, etc.) for any purpose, done by the attacker.
\\ \hline
\vspace{0.2pt}
\textbf{Obtain Sensitive Information - Credentials ($f_{z=11}$)} & the act of obtaining any user/system credentials without authorization via any method (unauthorized access to files, databases, memory, etc.) for any purpose, done by the attacker.
\\ \hline
\vspace{0.2pt}
\textbf{Read From Memory ($f_{z=12}$)} & the act of unauthorized reading any information or data from memory for any purpose, done by the attacker.\\ \hline
\vspace{0.2pt}
 &\textit{NOTE 2} "Obtain Sensitive Information - Other Data" and "Read From Memory" share exactly the same MITRE technique.  However, since the action and the purpose might differ, they are considered as separate functionalities.\\
\hline
\vspace{0.2pt}
 \textbf{Read Files ($f_{z=13}$)} & the act of unauthorized reading any information including from files, done by the attacker.
 \\ \hline
 \vspace{0.2pt}
 & \textit{NOTE 4} This functionality and "Obtain Sensitive Information" (($f_{z=10}$) and ($f_{z=11}$))" are mainly using the "Read" action mentioned in the previous functionalities, hence they share mutual characteristics with each other in terms of common MITRE techniques.\\
 \hline
 \vspace{0.2pt}
\textbf{Memory Read (Memory Buffer Errors, Pointer Issues, Type Errors, etc.) ($f_{z=14}$)} & the act of any invalid reading from the memory (e.g., buffer, kernel, memory locations, pointers, etc.) leading to memory issues such as buffer-over read, memory crash, buffer overflow, etc., done by the attacker.
\\ \hline
\vspace{0.2pt}
\textbf{Restart Or Reboot ($f_{z=15}$)} & the act of crashing, shutting down, or rebooting any system often leading to a denial of services, done by the attacker.
\\ \hline
\vspace{0.2pt}
\textbf{Write To Existing File ($f_{z=16}$)} & the act of modifying the content of the existing file for any purpose, done by the attacker.\\ 
\hline

\end{tabular}
\caption{The functionalities and their definitions}
\label{Tab: func_desc}
\end{table*}
Figure \ref{fig: functionality_relat} illustrates the comprehensive overview of all functionalities along with their corresponding dependencies. In our analysis, we have identified two distinct types of dependencies: commonality and inheritance. The term "inheritance" denotes that a child's functionality inherits all characteristics of its parent functionality, in addition to possessing its own unique attributes. These characteristics primarily pertain to the shared threat actions and/or MITRE techniques associated with the functionalities. For instance, to execute the \textit{out-of-bound-read} functionality, denoted as "$f_{z=14}$: Memory Read (Memory Errors)," an attacker must first perform the \textit{read} action, which is also a part of the "$f_{z=12}$: Read From Memory" functionality. Examples of such actions include \textit{reading arbitrary memory} or \textit{reading kernel memory}.

On the other hand, the concept of commonality refers to the semantic similarity between two functionality classes. It implies that both classes can describe similar impacts or actions, albeit not necessarily identical ones. In instances where the commonality is strong, it indicates that the impact can be the same, and distinguishing between the two classes may be challenging. For instance, an attacker can both install an extension ($f_z=5$) and manipulate system configurations ($f_z=9$), or alternatively, read memory locations ($f_z=14$) and gain sensitive information ($f_z=10$). In such cases, the actions or impacts can be used interchangeably for both classes. Conversely, when the commonality is weak, a single action or impact may imply two distinct concepts. For instance, when an attacker gains sensitive information, it could refer to obtaining a list of system files ($f_z=10$) or obtaining plain-text passwords ($f_z=11$).

It is important to note that in the context of strong commonality, actions or impacts may denote the same concept but are expressed using different terminologies, while in the case of weak commonality, actions or impacts refer to distinct concepts while being described using the same language. Based on the provided table, it is evident that types 13 and 14 inherit from type 12 (inheritance), and they also exhibit a strong commonality with types 10 and 11. Similarly, types 10 and 12 demonstrate a strong commonality, while types 5 and 9 also exhibit a strong commonality. Additionally, types 10 and 11 showcase a weak commonality. Given these dependencies, it becomes crucial to consider them during corpus generation, dataset creation, as well as model training and evaluation processes.

In summary, if $f_z$ and $f_z'$ represent two functionalities and $R(f_z, f_z')$ denotes the dependency relationship between them, the aforementioned dependencies play a pivotal role in corpus generation, dataset creation, and the training and evaluation of models:
\newline

\noindent$R(f_z, f_{z'}) = \textit{Inher}$ IF $f_{z'}$ is the inheritor of $f_z$

\noindent$R(f_z, f_{z'}) = \textit{Strong}$ IF $f_z$ and $f_{z'}$ have strong commonality dependency

\noindent $R(f_z, f_{z'}) = \textit{Weak}$ IF $f_z$ and $f_{z'}$ have weak commonality dependency 

\noindent $R(f_{z'}, f_{z''}) = \textit{Strong}$ IF  $R(f_z, f_{z'}) = \textit{Inher}$ and $R(f_z, f_{z''}) = \textit{Strong}$.
\newline

Each of the preceding functionality refers to a specific malicious action carried out by an attacker in order to compromise a system. 
Specifying the dependencies helps in better understanding and processing the main concept of each functionality leading to strategic text extraction and annotation, discussed in the next section.

%% file: Problem_Definition.tex
\section{Problem Definition}\label{sec: problem_def}
The problem addressed in this research revolves around the classification of CVEs into specific functionalities, providing a more comprehensive understanding of the potential impacts associated with each vulnerability. The absence of a readily available labeled dataset poses a significant challenge. In addition, it is critical to understand the exact implication of threat actions based on the context to be able to predict the correct intention and accordingly the final functionality. Thus, the research focuses on developing an approach that combines semantic role labeling, frequent verb, and object extraction from CVE reports, and a novel classifier design on top of the domain-specific language model to generate a labeled dataset and build an accurate classifier capable of assigning CVEs to their respective functionalities based on contextual information. In short, we aim at answering the following three research questions:\\

\textbf{RQ.1}:
How can  semantic role labeling techniques be effectively employed to generate a labeled dataset in the cybersecurity domain for different tasks?

\textbf{RQ.2}:
To what extent does the utilization of context-oriented classification model design enhance the performance and robustness of the classification approach?

\textbf{RQ.3}:
What are the strengths and limitations of the proposed methodology for the functionality-based classification of CVEs, and how can it be improved using state-of-the-art language models such as ChatGPT \cite{liu2023summary}?

%% file: DataAssessment.tex
\section{Data Extraction, Annotation, and Assessment}\label{sec: data_collection}
\begin{figure}[ht]
  \centering
  \includegraphics[width=9cm]{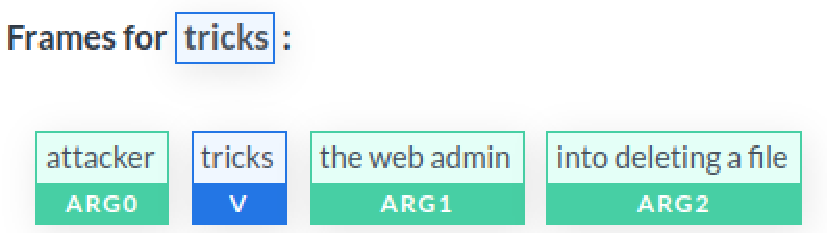}
  \caption{SRL breaks down a text into words or phrases as arguments and returns their semantic role in the sentence.}
  \label{fig: SRL_exmpl}
\end{figure}

SRL is a powerful NLP technique that identifies and categorizes the semantic roles of words or phrases in a sentence. SRL establishes the relationships between predicates (verbs) and their associated arguments (subjects, objects, etc.), providing a structured representation of syntactic and semantic properties.
SRL models possess the ability to analyze text by segmenting it into sentences based on identified verbs \cite{kambhatla2013systems}. These models can accurately identify the words or phrases associated with those verbs and assign them specific semantic roles within their respective sentences. For example, as demonstrated in Fig \ref{fig: SRL_exmpl}, using a pre-trained SRL model~\cite{allennlp-srl}, the sentence "attacker tricks the web admin into deleting a file" can be divided into two sentences: "attacker tricks the web admin to delete a file" and "the web admin deletes a file." In the first sentence, the SRL model identifies "attacker" as the subject (ARG0), "tricks" as the verb (V), "the web admin" as the object (ARG1), and "to delete a file" as an attribute (ARG2) of the object \cite{johansson2008dependency}.

Within this context, our approach involves the collection of SVOs corresponding to each functionality from the CVE descriptions. To achieve this, we leverage the AllenNLP Semantic Role Labeling (SRL) tool, which is a pre-trained model specifically designed for performing semantic role labeling on text data. This tool takes a sentence as input and generates the corresponding semantic roles for each word or phrase in the sentence.

The functionalities can be categorized into two distinct groups. The first group pertains to "actions" that precisely delineate the specific exploitation. For example, functionalities such as "read file," "create an account," and "install app" explicitly denote the exact actions an attacker would undertake subsequent to exploiting the vulnerability. On the other hand, the second group of functionalities characterizes the "impact" of the exploit rather than the "action" itself, focusing on the ultimate consequences of the attack. This category encompasses functionalities that describe memory errors, including buffer errors, for instance.

To extract statements that describe action functionalities, we represent each functionality with a set of SVOs in the form ${<Subject> <Verb> <Object>}$. The $<Subject>$ component represents any type of adversary, such as an attacker, hacker, or unauthorized user. The $<Verb>$ component refers to the action being performed, encompassing verbs such as "read," "write," "modify," and others. Lastly, the $<Object>$ component encompasses various cyber objects that are involved in the action, such as files, accounts, and so forth. This representation can be expressed as an SVO triplet, denoted as $SVO = <ARG0> <V> <ARG1>$.

Let's define an argument dictionary consisting of subjects (ARG0) $D_{subj}$, verbs (V) $D_{verb}$, and objects (ARG1) $D_{obj}$ that have been manually extracted and annotated from CVE reports.
If $D_{subj}=$\{attacker, adversary, hacker, unauthorized user, unauthenticated user\} is a constant dictionary of particular malicious actors, let $f_{z}$ denote the functionality with index $z$, and $V_{z} \in D_{verb}$ and $O_{z} \in D_{obj}$ be the set of verbs (V) and objects (ARG1) associated with functionality $f_{z}$. Both of these sets are extracted based on the relevancy to each functionality and the commonality of words in CVE reports based on expert knowledge. Therefore, given the list of annotated verbs and objects, for each functionality $f_{z}$, we utilize AllenNLP to extract every possible statement in terms of SVOs from CVE reports that contain $verbs \in V_z$ and $Objects \in O_z$ to generate the initially labeled dataset. Thus, we represent each functionality $f_{z}$ as the union of extracted SVOs as follows:

\begin{equation}
\footnotesize
    S(f_{z}) = \bigcup{}{}{S^z_iV^z_jO^z_k}\textit{ where } S^z_i \in D_{subj},V^z_j\in D_{verb}, O^z_k \in D_{obj}
    \label{eq: svo1}
\end{equation}
\FloatBarrier
\begin{table}[!ht]
    \centering
    \scriptsize

    \begin{tabular}{|p{0.25\textwidth}|p{0.6\textwidth}|}
    \hline
    
        \Centering\textbf{$f_z$} & \Centering\textbf{SVOs} \\ \hline \hline
 Create Account & 
        \par \textit{- remote attackers create new accounts}
        \par \textit{- unauthenticated users create accounts with arbitrary roles} 
        \\
 \hline
 Read Files &  
 \par \textit{- remote attacker read arbitrary files}
 \par \textit{- attackers view arbitrary files on the system} 
 \\ 
 \hline
 Change Ownership & 
 \par \textit{- remote attackers modify permission field} 
 \par \textit{- unauthenticated user changes the ownership of the files}\\ 
 \hline
 Install App &
 \par \textit{- unauthenticated, remote attacker install additional jee applications}
 \par \textit{- attacker place a malicious dll file}
 \\ \hline
    \end{tabular}
    \caption{Example of extracted SVOs for four functionalities}
    \label{tab: SVO_example}
\end{table}
\FloatBarrier
In Eq. \ref{eq: svo1},$z \in \{1,..,16\}$ is denoted as the functionality index. Additionally, $S^z_i$, $V^z_j$, and $O^z_k$ correspond to the subject, verb, and object associated with functionality $f_{z}$, respectively. $S^z_iV^z_jO^z_k$ also represents an SVO extracted by AllenNLP.

Table \ref{tab: SVO_example} shows a few examples of SVOs extracted for four functionalities. 
Such statements provide a plain and simple representation of functionalities with respect to particular actions and objects. 
For example, SVOs such as \textit{remote attacker read arbitrary files} which represent functionality "Read File" involving annotated verbs and objects  frequently appear in CVE reports that clearly describe any form of file (arbitrary, dll, txt, etc.) reading (read, open, access, etc.) in CVE reports that directly reflect the attacker's main action. 
However, this form of simple SVO extraction may not adequately describe impact functionalities or may not be quite typical for all functionalities. 
For example, both "\textit{attackers read memory}" and "\textit{attackers read memory that causes buffer over-read condition}" phrases provide actions addressing similar behavior. The former conforms to the simple format of SVOs we discussed above, whereas the latter is a bit different and provides more granular information regarding the impact of "buffer over-read" contrition. 
This form of threat representation is quite prevalent in the CVE report format that demonstrates a primary action or a fault in the system followed by an impact as {\{$<ACTION/FAULT> CAUSES <IMPACT>\}$. We refer to this as a causal link in which, an action causes or leads to an impact, which is important for describing impact functionalities \cite{gianvecchio2019closing}. In addition, defining such links can help to differentiate one class from another one. For example, reading memory by attacker refers to the $f_{6}=\text{\textit{Read From Memory}}$, but when this action causes buffer issues, it infers the $f_{9}$ known as \textit{Memory Read (Memory Buffer Errors)}.

We establish new sets of verbs and objects to extract SVOs related to impact functionalities, similar to the earlier approach for extracting SVOs associated with action functionalities.
Let $D'_{verb}=\text{\{cause, lead, result\}}$ and $D'_{Obj}$ represent another set of annotated verbs and objects corresponding to causal link, respectively. 

If $f_{z}$ represents the functionality with index $z$, we define a new set of objects $O'_{z} \in D'_{obj}$ associated with $f_{z}$. Likewise, for those classes that have this causal link, we extract the SVOs utilizing the defined rules. Similar to Eq. \ref{eq: svo1},  we represent each functionality $f_{z}$ as the union of extracted SVOs as follows:
\begin{equation}
\footnotesize
    S'(f_{z}) = \bigcup{}{}{S'^z_iV'^z_jO'^z_k}\textit{ where } S'^z_i \in D_{subj},V'^z_j\in D'_{verb}, O'^z_k \in D'_{obj}
    \label{eq: svo2}
\end{equation}

Therefore, every functionality $f_{z}$ is represented by the union of two sets of SVOs as $S(f_{z}) \cup S'(f_{z})$.

The interpretation and semantic meaning of a threat action is context-dependent, highlighting the importance of analyzing text within its surrounding context. This is especially crucial when classifying actions within a CVE description to functionalities. CVE descriptions contain detailed information about attacks or vulnerabilities, including specific threat actions performed by attackers. However, understanding the true implications and purpose of these actions requires considering the broader context of the description, along with the extracted phrases using SRL. 
The context provides insights into the attacker's motivations, techniques, and potential outcomes of the attack. For example, a threat action (the content) such as "bypassing authentication" may have different implications depending on the context. In one CVE description, it might refer to an attacker gaining unauthorized access to a system by circumventing authentication mechanisms. In another context, it could indicate an attacker exploiting a vulnerability to undermine or bypass the established authentication processes.
By considering the threat actions within the entire CVE description or generally, the document which the threat action is extracted from, analysts can precisely learn the corresponding functionality associated with each action. This holistic approach enables a deeper understanding of the attack techniques employed and the potential impact on the targeted system or network.

Along with the 110K CVE descriptions we analyzed to extract the contents and contexts, we  the AllenNLP on external resources, such as security advisories and security reports from NIST, NVD, MITRE, and other vendors such as Microsoft, RedHat, Apple, etc., to extract 10,000 content-context pairs. In addition, we have also annotated  1,098 SVOs manually relying on expert knowledge. We took the 75\% of data for training and the remaining 25\% for testing called $TS1$. In addition to the 25\% testing dataset, we have also manually labeled 494 CVEs by extracting their content-context pairs as another testing dataset, called $TS2$. In this dataset, contents are one or more SVOs extracted from CVE descriptions and the context is the entire CVE description.

It is important to emphasize that our goal is not to exhaustively capture every possible representation. Instead, we gather a significant amount of data and strategically train the model to understand the relationship between the extracted data and unseen text. This mechanism is instrumental in highlighting the essential content of the target functionality and capturing the surrounding context. By capturing the semantic association between content and context, our approach enables the identification of more complex actions and implicit patterns that cannot be solely derived from the content itself.

%% file: Model_Design.tex
\section{Methodology}\label{sec: cve2func_meth}
We retrieved a list of SVOs for each functionality in the previous section, with each SVO associated with a CVE report. Our goal is to predict the functionality of a certain action (e.g., \textit{read arbitrary file}) based on the surrounding context. 

We use SecureBERT's capacity to obtain the relationship between two phrases and leverage it to improve the text classification with document-level contextual information. Therefore, we develop a model that takes two inputs, "content" and "context," and returns the content's corresponding functionality.
In essence, the term "content" refers to a typically short text that describes a particular action (in this case, functionality) with no or minimal noise. Meanwhile, "context" implies a longer text which includes the content as well as additional discussions about related or similar notions.

As model input, we concatenate the content $X$ and context $D$ into a text sequence $[<[\textit{CLS}]> X <SEP> D <SEP>]$. Then, in a mini-batch, pad each text sequence to M tokens, where M is the batch's maximum length. After that, the vector [\textit{CLS}] is fed into a single-layer neural network with N output neurons, where N denotes the total number of functionalities. We begin with a pre-trained SecureBERT model and fine-tune it using cross-entropy loss.
To develop such a model, we would first establish a dataset in which each sample is a text pair, and then design the classification layer utilizing SecureBERT to classify the text pair into a functionality.\\
\newline
To train the model,  SVOs must be structured in a specified format and paired with another relevant text. As previously noted, the objective is to build a model that can predict the functionality corresponding to content (SVO) within a context. 
In other words, the same action may address multiple functionalities; thus, we aim to train a model that can identify the correlation between a particular action (content) and the longer text (context) which includes more information about the action to deliver the correct functionality to which the action refers.
Hence, it is crucial to strategically pair the SVOs with a relevant text, in order to maximize training performance.

To capture different types of information, we link each SVO with three types of pairs in different stages for creating the training dataset. It is worth noting that, for each functionality in all three stages, we generate a maximum $\tau_x$ number of pairs to avoid oversampling and the imbalanced data problem.
\newline

\noindent \textbf{First}, for each $f_z$, we generate $\tau_1$ number of samples by randomly pairing the SVOs belong to $f_z$. This helps the model learn threat actions' paradigmatic relation by observing different SVO representations associated with functionality. 

\noindent\textit{NOTE}:  Causal SVOs can be used as is in the content part (e.g., \textit{attacker cause buffer overread condition}). 
However, for $f_{12}$ (Memory Read) and $f_{16}$ (Memory Modification), the CVE description of the non-causal SVO must contain the causal objects to avoid class conflicts. For example, "\textit{attackers read memory}" can be associated with both $f_{12}$ (Read From Memory) and $f_{13}$ (Memory Read). This SVO would get $f_9$ label if the CVE description associated with $f_{13}$ contains an object in $o'_j$. In other words, we look for predefined objects defined for "Memory Read" within the CVE description and if found, the SVO "\textit{attackers read memory}" will be assigned to $f_9$, otherwise, it will be labeled as $f_{12}$.
\newline

\noindent \textbf{Second}, for each $f_z$, we generate $\tau_2$ number of samples by randomly pairing the SVOs (content) belonging to $f_z$ and a positive sentence (context) in the manually created dataset. Positive refers to the sentence we labeled as $f_z$ in the manual dataset. This helps the model to learn the relevancy of the content within the context of an extended vocabulary.
\newline

\noindent \textbf{Third}, for each $f_z$, we generate $\tau_3$ number of samples by pairing the SVOs (content) with the corresponding full CVE description (context). This helps the model learn the content of functionality within a noisy context. 
\newline

\begin{figure}[!ht]
\center
\includegraphics[width=10cm]{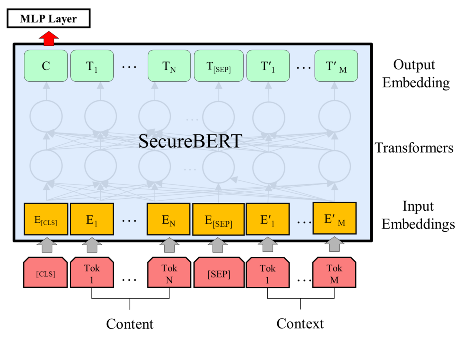}
\caption{The model architecture for classifying CVEs to functionalities}
\label{fig: SVO2Func_model}
\end{figure}

Fig. \ref{fig: SVO2Func_model} shows the classification model using SecureBERT. Unlike the standard way of text classification which models typically take only one input, our model is designed to take two inputs separated by a special token [\textit{SEP}] due to the specific characteristics of this classification problem. 
First, the second input pair, or context, may provide unseen yet relevant information about the extracted content (SVO) that the defined rules cannot capture. This creates a semantic link between the content and the unseen information, which greatly improves the model's learning and helps in better generalization.
In addition, functionalities are sometimes interdependent, and SVOs require associated context in order to be identified correctly. In other words, two functionalities may share the same SVO and the correct functionality for that SVO cannot be recognized unless it is assessed inside the context.
For example, Table \ref{tab: SVO_example} represents two SVOs as contents extracted from the given context. The statements "\textit{attackers read arbitrary kernel memory}" and "\textit{attackers read kernel memory}" represent both "Read From Memory" and "Memory Read (Memory Errors)" if used individually. However, when such content appears with the statement "\textit{cause a denial of service}" within the context, it clearly refers to the "Memory Read (Memory Errors)" as the corresponding functionality for the SVO.
In the meantime, the context describes the underlying weakness, which is improper access check by stating "\textit{does not perform certain required \textit{access\_ok} checks}", which leads to the functionality "Memory Read (Memory Errors)". Therefore, the model can establish a semantic link between the provided weakness and the functionality during the training, which can be used as the potential indicator for predicting unseen samples in the future.
\begin{table}[h]
\scriptsize
\centering
\begin{tabular}{|p{12cm}|} 
\hline
\multicolumn{1}{|c|}{\textbf{Contents}}\\ 
\hline
- attackers read arbitrary kernel memory\\ 
- attackers read kernel memory\\
\hline

\multicolumn{1}{|c|}{\textbf{Context}}\\ 
\hline
\textbf Linux kernel \textcolor[rgb]{0.95,0.31,0.41}{does not perform certain required \textit{access\_ok} checks}, which allows \textbf{attackers} to \textbf{read arbitrary kernel memory} on 64-bit systems and \textcolor[rgb]{0.95,0.31,0.41}{cause a denial of service} and possibly \textbf{read kernel memory on 32-bit systems}.\\
\hline
\end{tabular}
\caption{An example of contents and context as two inputs to the classification model.}
\label{tab: SVO_example}
\end{table}

%% file: Evaluation.tex
\section{Evaluation}\label{sec: eval}
For this multiclass classification task, we use Binary Cross Entropy (BCE) as the loss function. We use $2$ epochs with the Adam optimizer and an initial learning rate of $1e-5$. The model returns $16$ classification scores for each input pair corresponding to each functionality, with the highest score considered the model's final prediction. 
Table \ref{fig: cve2func_eval} shows the performance evaluation of the model on both testing datasets. According to the results, our proposed model shows high performance in predicting the correct functionalities corresponding to each input content within the given context, in both $TS1$ and $TS2$ testing datasets. 

\begin{table}[!ht]
\centering

\begin{tabular}{|l|l|l|} 
\hline
\textbf{\textbf{\textbf{\textbf{Metric}}}} & $TS1$ & $TS2$                     \\ 
\hline
Accuracy              &  0.981           &   0.983        \\
\hdashline
Precision~(Micro)     &  0.981           &   0.983        \\
Recall~(Micro)        &  0.981           &   0.982        \\
F1-Score~(Micro)      &  0.981           &   0.983        \\
\hdashline
Precision~(Macro)     &  0.947           &   0.979        \\
Recall~(Macro)        &  0.965           &   0.976        \\
F1-Score~(Macro)      &  0.954           &   0.975        \\
\hline
\end{tabular}
\caption{The performance of the CVE to functionality classification model using all testing datasets.}
\label{fig: cve2func_eval}
\end{table}
\begin{figure}[!ht]
 \center
 \includegraphics[width=\textwidth]{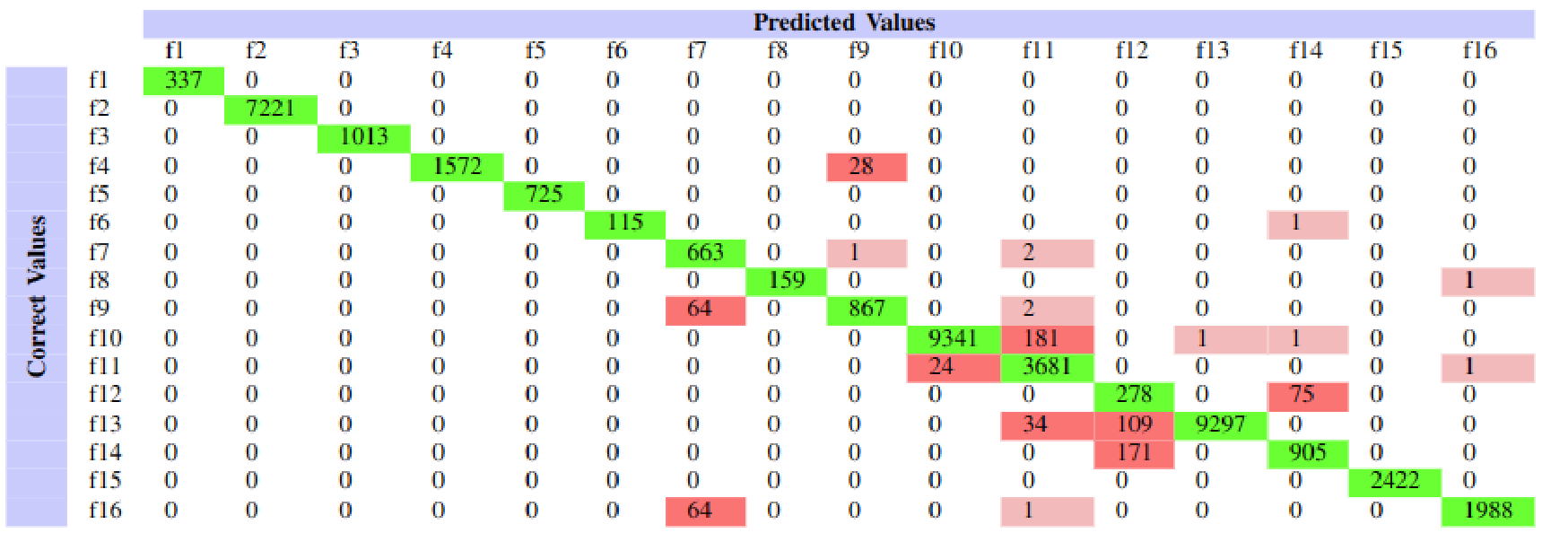}
\caption{The confusion matrix of threat actions to functionality classification model using all testing datasets (\textit{TS1 dataset}).}
\label{tab: cve2func_all_cm1}
\end{figure}

\begin{figure}[!ht]
 \center
  \includegraphics[width=11cm]{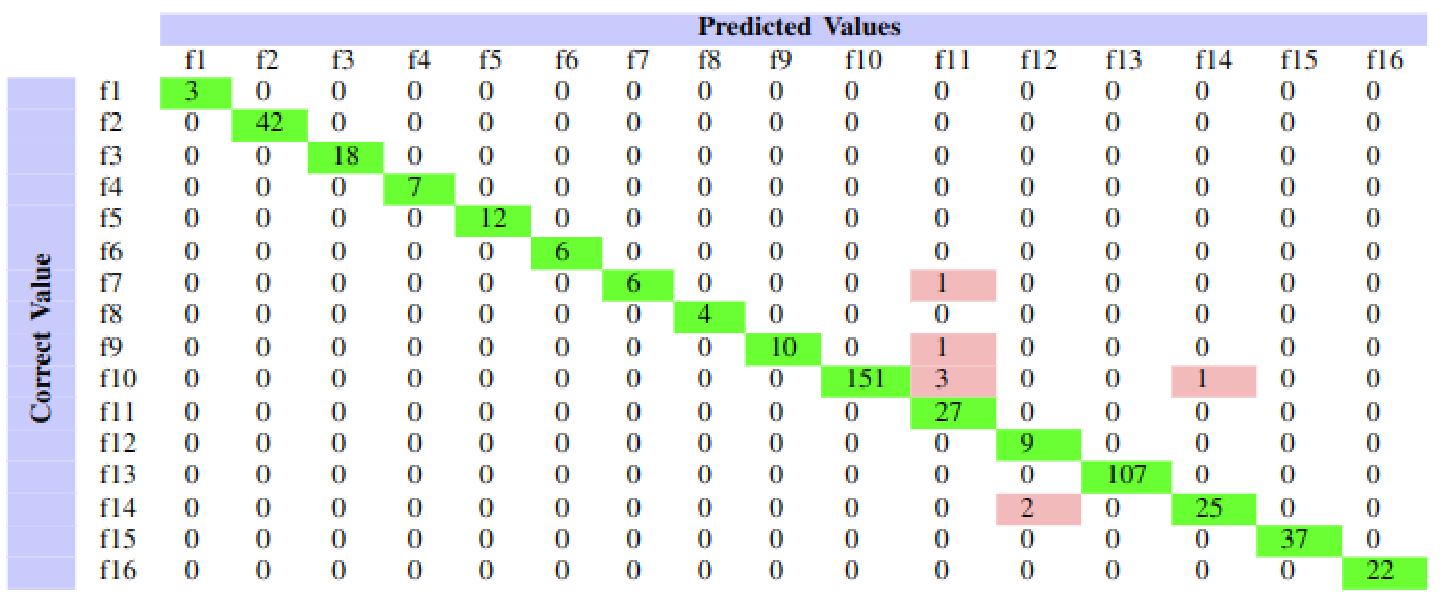}
\caption{The confusion matrix of threat actions to functionality classification model using 494 pairs of content and CVE descriptions as context (\textit{TS2 dataset}).}
\label{tab: cve2func_cve_cm2}
\end{figure}

Based on the confusion matrices depicted in Fig. \ref{tab: cve2func_all_cm1} and \ref{tab: cve2func_cve_cm2}, the model shows high performance in predicting the correct functionality despite the existing imbalance data problem in our dataset.
However, there is some confusion, particularly in a few classes that share some sort of dependencies as weak/strong commonality or inheritance. Functionality \textit{Read From Memory} ($f_{12}$), \textit{Read Files} ($f_{13}$),  and \textit{Memory Read (Memory Errors)} ($f_{14}$) which share inheritance dependencies,  return the highest confusion in the prediction. This confusion mostly happens in dataset $TS1$ in which, input pairs may share ambiguous meanings with multiple intentions. 
For example, the combination of some texts such as "\textit{local user read arbitrary kernel memory location.}", "\textit{local users, including low integrity processes, read and write to arbitrary memory locations.}", "\textit{local user read arbitrary memory}" to generate content-context pairs can refer to any "Read" action from the memory. Such texts as context if provided with a context potentially about the impact or further intuitions can lead to better prediction. Similarly, \textit{Obtain Sensitive Information - Other Data
} ($f_{10}$) and \textit{Obtain Sensitive Information - Credentials} ($f_{13}$) can also share similar properties in some inputs leading to confusion. For example, \textit{sensitive data} or "sensitive information"  in  "\textit{remote attacker obtain sensitive data}" or "\textit{local attacker obtain sensitive information}" may refer to "credentials" or other types of information such as "directory/file names" or "browser history", etc., so they must be followed by an informative context such as "\textit{XXX could allow a local attacker to obtain sensitive information, caused by plain text \textbf{user account passwords} potentially being stored in the browser's application command history. By accessing browser history, an attacker could exploit this vulnerability to obtain other \textbf{user accounts' passwords}.}" for better prediction.
The evaluation on dataset $TS2$ which indeed contains precise context (entire CVE descriptions) that is depicted in Fig. \ref{tab: cve2func_cve_cm2} shows higher performance in functionalities whose characteristics have some types of dependencies.


In real-world applications, the "content" may not always be available. In practice, cybersecurity researchers or software vendors often only have access to a security report (e.g., a CVE description) and intend to classify it to functionalities. Without requiring the second input, our proposed model can take such security reports and return the corresponding scores to each of the sixteen functionalities. In fact, the trained model has been processed in order to capture the contextual representation of the input document in accordance with any of the functionalities. 
It is worth noting that, delivering all functionalities associated with a document, that can include more than one functionality, implies a multi-label multiclass classification approach. However, in the absence of a standard multi-label dataset, it is not feasible to train the model in such a way. Thus, as a proof-of-concept, we evaluated the model on 66 CVE descriptions that are mapped to one or more functionalities by MITRE in the guideline (detailed results: \footnote{\url{https://github.com/ehsanaghaei/CVE2TTP/blob/main/CVE2FUNC_appendix.pdf}}).
In other words, if a CVE is associated with \textit{m} functionalities, we show at least \textit{m} top predictions. If all returned predictions exactly match the ground truth, we do not provide any further prediction, and hence, $K=m$. However, if all the correct classes are not covered in the top \textit{m} predictions, we show another \textit{m'} prediction until all correct classes are covered, and therefore $K = m+m'$. Based on the results, in 58 out of 66 data samples, the top K model predictions are equal to the number of correct classes ($K=m$ and $m'=0$) implying an $87.88\%$ overall hit rate. In 6 data samples, the model returns one extra prediction to cover all correct classes (($K=m + m'$ and $m'=1$) which represents a cumulative $96.97\%$ hit rate with one false positive. Overall, the model shows a $100\%$ hit rate in the top 5 ($K=5$) predictions for all test data.

The promising performance confirms that the CVE language is effectively assessed in the dataset creation and that the model appropriately captured the textual features and semantic relationship between the text and the functionalities during training.\\

\noindent\textbf{Comparison with ChatGPT}\\
While recent advancements in large language models have shown satisfactory results in handling general text analysis tasks, they often fall short when it comes to specialized cybersecurity tasks that require precise domain knowledge. In contrast to ChatGPT, our model employs an ongoing process of learning and improvement to consistently achieve a high level of accuracy in vulnerability assessment through automated classification of threat actions into TTPs. To evaluate ChatGPT's capacities to classify CVEs to MITRE ATT\&CK techniques, we tasked it with predicting 20 CVEs using "Prompt: Find the associated MITRE ATT\&CK technique for <CEV Description>", each of which had previously been manually classified to their respective techniques by MITRE experts.
Table \ref{tab: cgpt} shows the predicted and ground truth for each CVE example demonstrating $80\%$ failure in correctly predicting CVEs to the corresponding MITRE ATT\&CK techniques. The disparity in classification underscores the critical role of domain-specific knowledge. In the cybersecurity domain, where a nuanced understanding of attack patterns, vulnerabilities, and threat behaviors is paramount, relying solely on a general-purpose model like ChatGPT can lead to substantial misclassifications. A domain-specific model is finely tuned to capture intricate nuances and context-specific to the field, enabling more accurate and reliable classification of CVEs into relevant MITRE ATT\&CK techniques. This further emphasizes the significance of employing domain-specific tools that possess the expertise and contextual awareness necessary to make informed decisions and analyze potential security risks effectively.

\begin{table*}[!ht]
\centering
\begin{tabular}{|p{3cm}|p{5cm}|p{5cm}|}
\hline
\textbf{CVE ID} & \multicolumn{1}{c|}{\textbf{Predicted}}                                                                                                    & \multicolumn{1}{c|}{\textbf{Ground Truth}}                       \\ \hline
CVE-2019-15243  & Exploitation for Privilege Escalation (T1068)                                                                                              & Command and Scripting Interpreter(T1059)                         \\ \hline
CVE-2019-15976  & Bypass User Account Control (T1088)                                                                                                        & Exploitation for Privilege Escalation (T1068)                    \\ \hline
CVE-2019-15956  & Resource Hijacking (T1496)                                                                                                                 & Endpoint Denial of Service(T1499)                                \\ \hline
CVE-2020-3253   & Exploitation for Privilege Escalation (T1068)                                                                                              & Command and Scripting Interpreter(T1059)                         \\ \hline
CVE-2020-5331   & Credential Dumping (T1003)                                                                                                                 & Data from Local System (T1005)                                   \\ \hline
CVE-2020-3312   & {\color[HTML]{036400} \textbf{Data from Local System (T1005)}}                                                                             & Data from Local System (T1005)                                   \\ \hline
CVE-2019-1768   & Exploitation of Vulnerability (T1203)                                                                                                      & Stage Capabilities (T1608)                                       \\ \hline
CVE-2019-1724   & Session Hijacking (T1550.002)                                                                                                              & Remote Service Session Hijacking (T1563)                         \\ \hline
CVE-2019-1620   & Scripting (T1064)                                                                                                                          & Ingress Tool Transfer (T1105)                                    \\ \hline
CVE-2019-1886   & Endpoint Denial of Service: Service Exhaustion Flood (T1499.002)                                 & Service Stop (T1489)                                             \\ \hline
CVE-2019-1703   & {\color[HTML]{036400} \textbf{Endpoint Denial of Service: Service Exhaustion Flood (T1499.002)}} & Endpoint Denial of Service: Service Exhaustion Flood (T1499.002) \\ \hline
CVE-2020-3476   & File and Directory Permissions Modification (T1222)                                                                                        & Stored Data Manipulation (T1565.001)                             \\ \hline
CVE-2019-15974  & Spearphishing Link (T1192)                                                                                                                 & Data Manipulation: Stored Data Manipulation (T1565.002)          \\ \hline
CVE-2019-1876   & Multi-Stage Channels (T1104)                                                                                                               & Exploit Public-Facing Application (T1190)                        \\ \hline
CVE-2020-3237   & File Deletion (T1107)                                                                                                                      & Stored Data Manipulation(T1565.001)                              \\ \hline
CVE-2019-3707   & {\color[HTML]{036400} \textbf{Exploit Public-Facing Application (T1190)}}                                                                  & Exploit Public-Facing Application (T1190)                        \\ \hline
CVE-2019-18573  & Steal or Forge Kerberos Tickets (T1558.003)                                                                                                & Remote Service Session Hijacking (T1563)                         \\ \hline
CVE-2018-15784  & {\color[HTML]{036400} \textbf{Man-in-the-Middle (T1557)}}                                                                                  & Man-in-the-Middle (T1557)                                        \\ \hline
CVE-2020-5378   & Pre-OS Boot:  System Firmware (T1542.001)                                                                                                  & Pre-OS Boot:  System Firmware (T1542.001)                        \\ \hline
\end{tabular}

\caption{Examples of ChatGPT's CVE to technique classification.}
\label{tab: cgpt}
\end{table*}

%% file: RelatedWorks.tex
\section{Related Works}
Recently, substantial efforts have been dedicated to the development of text analytics tools aimed at extracting threat information from unstructured text. Additionally, the academic community has shown considerable interest in the MITRE ATT\&CK framework as a means to link other threat and vulnerability data to it.

Chen et al. \cite{chen2021threat} proposed a model to extract threat actions by using information retrieval techniques. To capture threat actions, this study uses word vectors, tagging, and filtering algorithms. The proposed solution automatically generates a key threat action list as the foundation of the ontology, uses a two-stage key threat action extraction technique, and uses word vector models for key threat extraction. This work labels tokens in a phrase with their grammatical word categories using part-of-speech tagging, but it does not maintain grammatical links between them, resulting in a limited semantic that is poorly usable for linking cybersecurity text. 

In \cite{zhang2021ex}, the authors present a mechanism for automatically extracting threat actions from APT reports and producing TTPs. Threat actions are extracted from APT reports using a BERT-BiLSTM-CRF-based extractor, and these extracted threat actions are then mapped to ontology to construct their related TTPs using TF-IDF. The actions, which include the subject, verb, and object, are extracted using EX-Action. Additionally, it offers a technique for extracting entity relations, which connect entities contextually and semantically. This approach has limited threat action extraction capabilities due to its over-reliance on semantic and part-of-speech analysis that fails to identify pronoun referents. In addition, this approach does not extend to linking vulnerability and threat information.

Ayoade \textit{et al}. \cite{ayoade2018automated} leveraged natural language processing techniques to extract attacker actions from $18,257$ threat report documents generated by different organizations and automatically classified them into standardized tactics and techniques. The lack of labeled data and non-standard report formats are the main challenge this paper addresses using the bias correction mechanism approach. In this work, text descriptions of reports are tokenized, the TF-IDF score for each word is calculated, and different bias correction mechanisms are applied to overcome non-standard format. As reported by this work, this approach shows a very low accuracy, close to 60\% in classifying threat information to ATT\&CK techniques when using commonly used threat reports such as APT notes and Symantec datasets as test data.

BRON \cite{hemberg2020linking}, is a comprehensive framework that combines multiple public sources of cyber threat and vulnerability information, namely MITRE's ATT\&CK MATRIX, CWE, CVE, and CAPEC. BRON maintains all entries and relationships while facilitating bidirectional path tracing. It utilizes attack patterns to establish connections between attack objectives, means, vulnerabilities, and the targeted software and hardware configurations. The authors conduct an inventory and analysis of BRON's sources to assess gaps between information on attacks and their targets. Additionally, they analyze BRON for any incidental information obtained during its mission. Despite the large database of threat information collected in this work, this dataset is unusable for training CVE-Technique classification models because the labeling is too general/abstract and does not lean toward CVE and MITRE techniques specifics.

Kuppa \textit{et al}. \cite{kuppa2021linking} propose using the MITRE ATT\&CK taxonomy to map CVEs to attack techniques. They introduce a Multi-Head Joint Embedding Neural Network model and unsupervised labeling to automate this process. Enriching CVEs with a knowledge base of mitigation strategies and attack scenarios improves understanding. The evaluation shows mapping many CVEs to ATT\&CK techniques, but limitations, including limited coverage of only 17 techniques and a small knowledge base, make the solution impractical.

Grigorescu \textit{et al}. \cite{grigorescu2022cve2att} addresses a standardized cyber-security knowledge database by annotating a dataset of CVEs with MITRE ATT\&CK techniques. Their paper presents models to automatically link CVEs to techniques using the text description from CVE metadata. They utilize classical machine learning models and BERT-based language models, addressing imbalanced training sets with data augmentation. The best model achieved an F1-score of 47.84\%. However, the study's limitations include a small training set of only 1813 CVEs and mapping to a limited set of 31 MITRE techniques, along with a low performance, indicating a lack of generalization in their proposed model.

Ample \textit{et al}. \cite{ampel2021linking} aim to build a cybersecurity knowledge base for critical infrastructure defense. They propose the CVE Transformer (CVET) model to label CVEs with ten ATT\&CK tactics. The model utilizes fine-tuning and self-knowledge distillation with RoBERTa, achieving a 76.1\% F1-score in labeling CVEs. The study uses a CVE dataset from BRON\cite{hemberg2020linking}, which provides classifications into higher-level abstractions, including MITRE ATT\&CK techniques and tactics. However, the interconnected mappings between CVEs and tactics through CAPEC lack the necessary granularity, and mapping to tactics does not offer fine-grain classification for specific techniques related to CVEs.

%% file: Conclusion.tex
\section{Conclusions}
This research addresses a pivotal cybersecurity challenge, entailing the characterization of the causal linkage between vulnerability information derived from CVE descriptions and the potential threat actions and techniques delineated by the ATT\&CK TTP framework.

To surmount this challenge, we confront foundational obstacles arising from the limited availability of training datasets for CVE classification and the inherent semantic divergence between CVEs and the ATT\&CK TTP framework.

By harnessing Semantic Role Labeling (SRL) to extract threat actions from an extensive corpus of unstructured cybersecurity reports, which encompasses CVEs, and through the extensive refinement of a domain-specific language model tailored for cybersecurity, we establish a comprehensive and meticulously annotated dataset. This dataset serves as a potent resource for efficacious training and the precise classification of CVEs into corresponding attack techniques (TTPs).

Our rigorous evaluation of this methodology attests to its efficacy, yielding remarkable accuracy rates of approximately $98\%$, accompanied by F1-scores spanning from $95\%$ to $98\%$ in the precise classification of CVE threat actions with respect to their corresponding ATT\&CK techniques. Furthermore, we substantiate that the performance of the TTPpredictor surpasses that of state-of-the-art language model tools, including ChatGPT. Through numerous iterations involving sets of 100 randomly selected CVEs, TTPpredictor achieves an average accuracy of $93\%$, contrasting with ChatGPT's accuracy of $20\%$ (see Table~\ref{tab: cgpt}).

In conclusion, our research not only pushes the boundaries of cybersecurity classification but also emphasizes the crucial significance of domain-specific language models and semantic comprehension in conquering intricate obstacles within this field. Moving forward, our future endeavors involve delving into the classification of CVEs and threat reports for defensive actions, coupled with an assessment of their effectiveness. 

